\documentclass{article}
\usepackage{graphicx}

\newcommand{\mb}[1]{\mbox{\normalsize\boldmath $#1$}}

\newcommand{\eq}[1]{~{\rm (\ref{eq:#1})}}

\newcommand{\GeV}{\,{\rm GeV}}
\def\circa#1{\,\raise.3ex\hbox{$#1$\kern-.75em\lower1ex\hbox{$\sim$}}\,}
\newcommand{\fig}[1]{~{\rm \ref{fig:#1}}}

\def\Red  {}
\def\Black{}
\def\Blue {}
\newcommand{\NP}{Nucl. Phys.}
\newcommand{\PRL}{Phys. Rev. Lett.}
\newcommand{\PL}{Phys. Lett.}

\newcommand{\BR}{\hbox{BR}\,}

\newcommand{\eV}{\,{\rm eV}}
\newcommand{\meV}{\,{\rm meV}}

\newcommand{\TeV}{\,{\rm TeV}}

\def\circa#1{\,\raise.3ex\hbox{$#1$\kern-.75em\lower1ex\hbox{$\sim$}}\,}
\makeatletter

\def\art{\@ifnextchar[{\eart}{\oart}}
\def\eart[#1]#2#3#4#5#6{{\rm #2}, {#3 \bf #4} {\rm (#6) #5} #1}
\def\hepart[#1]#2{{\rm #2, #1}}
\newcommand{\oart}[5]{{\rm #1}, {#2 \bf #3} {\rm (#5) #4}}

\newcounter{alphaequation}[equation]
\def\thealphaequation{\theequation\hbox to
0.6em{\hfil\alph{alphaequation}\hfil}}
\def\eqnsystem#1{
\def\@eqnnum{{\rm (\thealphaequation)}}
\def\@@eqncr{\let\@tempa\relax \ifcase\@eqcnt \def\@tempa{& & &} \or
  \def\@tempa{& &}\or \def\@tempa{&}\fi\@tempa
  \if@eqnsw\@eqnnum\refstepcounter{alphaequation}\fi
\global\@eqnswtrue\global\@eqcnt=0\cr}
\refstepcounter{equation} \let\@currentlabel\theequation \def\@tempb{#1}
\ifx\@tempb\empty\else\label{#1}\fi
\refstepcounter{alphaequation}
\let\@currentlabel\thealphaequation
\global\@eqnswtrue\global\@eqcnt=0 \tabskip\@centering\let\\=\@eqncr
$$\halign to \displaywidth\bgroup \@eqnsel\hskip\@centering
$\displaystyle\tabskip\z@{##}$&\global\@eqcnt\@ne
\hskip2\arraycolsep\hfil${##}$\hfil& \global\@eqcnt\tw@\hskip2\arraycolsep
$\displaystyle\tabskip\z@{##}$\hfil
\tabskip\@centering&\llap{##}\tabskip\z@\cr}
\def\endeqnsystem{\@@eqncr\egroup$$\global\@ignoretrue} \makeatother

\font\tenrsfs=rsfs10
\font\sevenrsfs=rsfs7
\font\fiversfs=rsfs5
\newfam\rsfsfam
\textfont\rsfsfam=\tenrsfs
\scriptfont\rsfsfam=\sevenrsfs
\scriptscriptfont\rsfsfam=\fiversfs
\def\mathscr#1{{\fam\rsfsfam\relax#1}}
\def\Lag{\mathscr{L}}

\begin{document}\twocolumn[
\centerline{hep-ph/0210021 \hfill CERN--TH/2002--262\hfill IFUP--TH/2002--39}
\vspace{5mm}
\Black
\vspace{0.5cm}
\centerline{\LARGE\bf\Red Predictions of the most minimal see-saw model}

\medskip\bigskip\Black

\centerline{\large\bf
M. Raidal$^{1,2}$ and A.\ Strumia$^{1,3}$}\vspace{0.4cm}

\centerline{\em $^1$Theoretical Physics Division,
CERN, CH-1211 Gen\`eve 23, Switzerland }
\centerline{\em $^2$National Institute of Chemical Physics and Biophysics,
Tallinn 10143, Estonia}
\centerline{\em $^3$ Dipartimento di Fisica
dell'Universit\`a di Pisa and INFN, Italy}
 \vspace{3mm}

 \begin{quote}\Blue
We derive the most minimal see-saw texture from an
extra-dimensional dynamics.
It predicts $\theta_{13} = 0.078\pm0.015$ and
$m_{ee} = 2.6\pm0.4\,\meV$.  Assuming thermal leptogenesis, the sign of
the CP-phase measurable in neutrino oscillations, together with the
sign of baryon asymmetry, determines the order of heavy neutrino masses.
Unless heavy neutrinos are almost degenerate, successful leptogenesis
fixes the lightest mass. Depending on the sign of the neutrino CP-phase,
the supersymmetric version of the model with universal
soft terms at high scale predicts BR($\mu\to e \gamma$) or
BR($\tau\to \mu \gamma$), and gives a lower bound on the other process.
\Black
\end{quote}
]

\noindent
\subsubsection*{\em Introduction}
The minimal see-saw~\cite{seesaw} texture that allows to explain the solar and
atmospheric neutrino anomalies in terms of oscillations
contains two heavy singlet neutrinos $N_{\rm atm}$ and $N_{\rm sun}$
coupled as~\cite{FGY}
\begin{eqnarray}\label{eq:txt}
\Lag &=& \Lag_{\rm SM}  + \frac{M_{\rm atm}}{2} N_{\rm atm}^2 +
\frac{M_{\rm sun}}{2} N_{\rm sun}^2 \\
&&+\lambda_{\rm sun} \,HN_{\rm sun}\, (s L_e + c\,\,e^{-i\phi/2} L_\mu +
0 \,L_\tau) \nonumber  \\
&&+\lambda_{\rm atm} HN_{\rm atm} (0 L_e + s_{\rm atm} L_\mu +
c_{\rm atm} L_\tau) , \nonumber
\end{eqnarray}
where $M_i$, $\lambda_i$, $\phi$, $s$ and $s_{\rm atm}$ are free parameters.
We abbreviate $s_i = \sin\theta_i$, $c_i = \cos\theta_i$, $t_i =\tan\theta_i$.
Nothing can be removed from this `most minimal texture'  without
generating conflicts with present data.
The phase $\phi$ is the unique source of
CP-violation in the lepton sector~\cite{ehlr}.
Possible connection~\cite{Branco,mor} between the sign of the
observed baryon asymmetry of the universe and the CP-violation in neutrino
oscillations via $\phi$ was the original motivation
of the model~\cite{FGY,mor}.
This texture is predictive also if the zeros are replaced by
sufficiently small numbers.
% The small entry in the $N_{\rm sun}$ coupling must be $\ll1$,
% the small entry in the $N_{\rm atm}$ coupling must be
%$\ll\sqrt{\Delta m^2_{\rm sun}/\Delta m^2_{\rm atm}}$.

In this letter we motivate the model\eq{txt} and
show in detail how it can be
tested using low and high-energy observables.
After deriving predictions for neutrino experiments,
we clarify that the sign of the baryon asymmetry, together with the sign
of the neutrino CP-violation in oscillations, determines a discrete
ambiguity of the model: the order of $M_{\rm sun}$ and $M_{\rm atm}.$
Successful leptogenesis~\cite{fy}
 fixes the mass of the lightest heavy neutrino.
In the supersymmetric version of the model,
either BR($\mu\to e \gamma$) or BR($\tau\to \mu \gamma$) is predicted,
depending on the sign of the neutrino CP-phase.
The other process is a function of the heaviest singlet
neutrino mass only, and has a lower bound.

\bigskip

At first sight the texture\eq{txt} looks quite artificial:
e.g.\ we do not know  how
a U(1) flavour symmetry could justify it.
However,\eq{txt} can be easily obtained from extra-dimensional models.
Following~\cite{HS}
we consider a 5-dimensional fermion $\Psi(x)$ in presence of a
domain wall $\varphi(x_5)$.
The system is described by the action
$$S = \int d^5x\, \bar{\Psi} [i\partial\hspace{-1.2ex}/\,  +
\lambda \varphi(x_5) - m]\Psi .$$
The Kaluza-Klein spectrum of $\Psi$ contains
a massless chiral mode localized around $x_5=x_5^*$,
where $\lambda\varphi(x_5^*) =m$.
When $\varphi(x_5)$ can be approximated with a linear function,
the chiral zero mode has a Gaussian profile in the extra dimension
with $\lambda$-dependent width.
Assuming that the Higgs $H$ is not localized,
small Yukawa couplings between $N$ and $L$ are naturally given by
a small overlap between their wave functions as depicted in the
figure below
$$\includegraphics{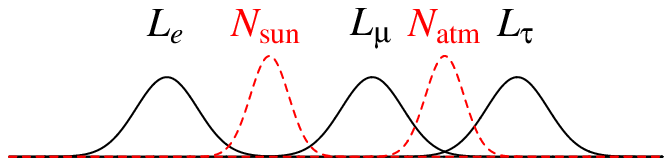}$$
This setup naturally generates
the desired matrix of Yukawa couplings
$$\bordermatrix{&L_e & L_\mu & L_\tau\cr
 N_{\rm sun}  & {\cal O}(\epsilon^2)  &
{\cal O}(\epsilon^2)& {\cal O}(\epsilon^{11.})\cr
N_{\rm atm} & {\cal O}(\epsilon^{14.}) &
{\cal O}(\epsilon) & {\cal O}(\epsilon)}$$
(where $\epsilon$ is a free parameter)
% 11. in place of 11 signals that the true number is 11.something
and suppresses $N_{\rm sun} N_{\rm atm}$  mixing mass terms.
While we do not gain any new insight proceeding along this route,
we are motivated to study the implications of the model.

 \begin{figure}[t]
 $$\includegraphics[width=8.3cm]{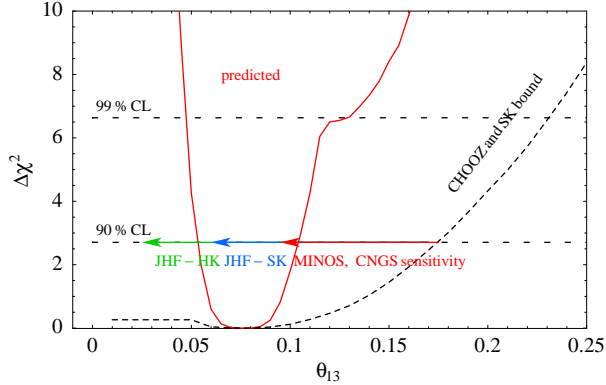}$$
 \caption{\em Prediction for $\theta_{13}$ 
and present bound from CHOOZ and SK.
The arrows indicate the expected sensitivity of future experiments.
 \label{fig:theta13}}
 \end{figure}

\subsubsection*{\em Neutrinos}
The model\eq{txt} predicts the following
Majorana mass matrix for the light neutrinos:
$$m_\nu = m_{\rm atm}
\pmatrix{ \epsilon s^2 & \epsilon sc e^{-i\phi/2} &
0 \cr \epsilon sc e^{-i\phi/2} & s^2_{\rm atm} +
\epsilon c^2 e^{-i\phi} & s_{\rm atm} c_{\rm atm}\cr 0 &
s_{\rm atm} c_{\rm atm} & c_{\rm atm}^2},$$
where $$m_{\rm atm} = \frac{\lambda_{\rm atm}^2 v}{M_{\rm atm}},\qquad
\epsilon =\frac{ \lambda_{\rm sun}^2/M_{\rm sun}}
{\lambda_{\rm atm}^2/M_{\rm atm}}.$$
$N_{\rm atm}$ plays the r\^ole of `dominant right-handed neutrino'~\cite{King}.
Neutrinos have a hierarchical mass spectrum and the lightest neutrino is
massless\footnote{An alternative minimal
texture, where $N_{\rm sun}$ couples to $L_\tau$ rather than to $L_\mu$,
is equally acceptable. Its predictions concerning neutrinos
can be obtained exchanging in the equations below
$\theta_{23}\leftrightarrow \pi/2 - \theta_{23}$.
At the moment atmospheric data do not distinguish between them.

If the singlet neutrinos have a pseudo-Dirac mass
term $M\, N_{\rm sun}N_{\rm atm}$,
rather than the masses of eq.\eq{txt},
one gets light neutrinos with inverted mass hierarchy.
Without fine-tuning its parameters,
the resulting texture predicts $\theta_{12}\approx \pi/4$
which is strongly disfavoured by the present data.}.
At leading order in $\epsilon$, the oscillation parameters are
$$
\Delta m^2_{\rm atm} = m_{\rm atm}^2 >0,\qquad
\Delta m^2_{\rm sun} = R {\Delta m^2_{\rm atm}},$$
with $R= \epsilon^2(s^2 + c^2 c_{\rm atm}^2)^2$.
We define $m_{\rm sun}\equiv \sqrt{\Delta m^2_{\rm sun}}$.
The mixing angles in the standard notation are
$$
\theta_{23} = \theta_{\rm atm},\qquad
\theta_{13} = \epsilon sc s_{\rm atm},\qquad
\tan\theta_{12} = \frac{s}{c c_{\rm atm}}.$$
The neutrino mixing matrix $V$
relating the mass eigenstates $\nu_i$
to the flavour eigenstates, $\nu_\ell=V_{\ell i} \nu_i$, is
\begin{eqnarray*}
 V &=&\hbox{diag}\,(1,e^{-i\frac{\phi}{2}},e^{-i\frac{\phi}{2}})\cdot
R_{23}(\theta_{23}) \cdot
\hbox{diag}\,(1,  e^{i \phi},1) \\
&&\cdot R_{13}(\theta_{13}) \cdot
R_{12}(\theta_{12}) \cdot\hbox{diag}\,(1,1,e^{i\frac{\phi}{2}}),
\end{eqnarray*}
where $R_{ij}(\theta_{ij})$ represents a
rotation by $\theta_{ij}$ in the $ij$ plane.
The first phase matrix in $V$ is unphysical and can be absorbed into
the phases of $(L_e,L_\mu,L_\tau)$.
The last phase matrix contains a
practically unmeasurable Majorana phase.
The phase matrix in the middle determines that the CP-violating phase
in oscillations (observable in the planned experiments) is exactly the phase $\phi$
in\eq{txt}.
%  The vacuum oscillations neutrino transition probabilities
%  after a  path-length $L$ is
%  $$P = | \exp(-iL m_\nu^\dagger m_\nu/2E_\nu)|^2$$
A `positive' phase, $0<\phi <\pi$ induces
$P(\nu_e\to \nu_\mu) > P(\nu_\mu\to \nu_e) = P(\bar{\nu}_e \to \bar{\nu}_\mu)$
in vacuum oscillations with
baseline $L < 2\pi E_\nu/\Delta m^2_{\rm sun}$.
% (this condition is satisfied in forthcoming beam experiments).

%  and
%  $$R_{13}(\theta_{13},\phi) =\pmatrix{\cos\theta_{13} & 0 &e^{i\phi} \sin\theta_{13} \cr
%  0 &1 &0 \cr -e^{-i\phi} \sin\theta_{13} &0&\cos\theta_{13}}.$$

\begin{figure*}[t]
 $$\includegraphics[width=15cm]{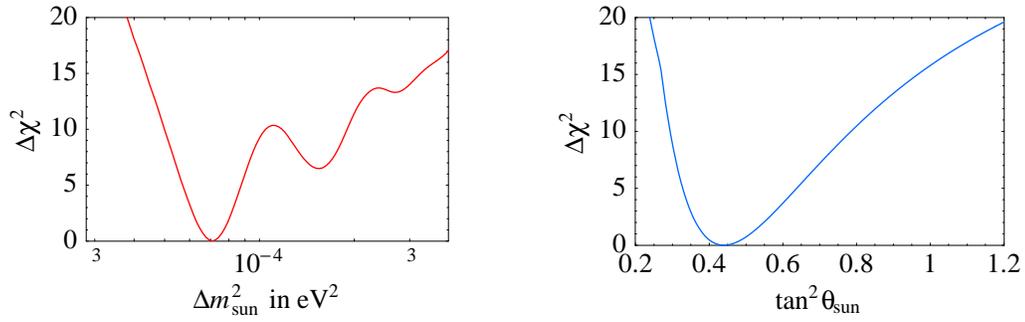}$$
\caption{\em
Best-fit values of the solar oscillation parameters, as obtained
 from solar and KamLAND data.
\label{fig:solarfit}}
\end{figure*}

Therefore this model {\em predicts} (see also~\cite{FGY})
\begin{equation}\label{eq:preds}
\theta_{13}\simeq
\frac{\sqrt{R}}{2} \sin2\theta_{12}\,\tan\theta_{23},\;\;\;
m_{ee} = m_{\rm sun}\sin^2 \theta_{12},
\end{equation}
where $m_{ee}$ is the $ee$ element of the neutrino mixing matrix
to be measured in neutrino-less double-beta ($0\nu2\beta$) decay
experiments. Present atmospheric neutrino data indicate
$\Delta m^2_{\rm atm}\approx 2.7~10^{-3}\eV^2$ and
$t_{23}^2\approx 1$ \cite{atm}. 
Solar and reactor data~\cite{sunexp,sunfit} indicate 
$\Delta m^2_{\rm sun}\approx 7.0~10^{-5}\eV^2$
and $\tan^2\theta_{12} \approx 0.45$
(another solution with a sligthly higher value of $\Delta m^2$
is somewhat disfavoured by data, see Fig.\fig{solarfit}).
Combining present atmospheric and solar data\footnote{
%We proceed as more precisely described in~\cite{0nu2beta}.
A function $\chi^2(p)$, extracted from a global up-to-date fit of solar and atmospheric data, contains the
present information
on the parameters $p=\{\theta_{\rm sun},
\theta_{\rm atm},\Delta m^2_{\rm sun},\Delta m^2_{\rm atm}\}$.
In Gaussian approximation,
the present information on a parameter $a=f(p)$ is then extracted computing
$\chi^2(a) = \min_{p:f(p) = a}\chi^2(p).$
$\chi^2(a)$ is `distributed as $\chi^2$ with 1 degree of freedom'.
In Bayesian inference this means
that the probability of different $a$ values is $\propto e^{-\chi^2(a)/2}$.
},
we obtain the predictions
\begin{equation}\label{eq:numpred}
\theta_{13} = 0.078\pm 0.015,\qquad
m_{ee} = 2.6\pm0.4 \,\meV.
\end{equation}
%  99\% CL predicted ranges (1 dof)
%  $$\theta_{13} = 0.03\div 0.13,\qquad
%  m_{ee} = 1.5\div 5 \,\meV.$$
% T
The predicted value of $m_{ee}$ is below the sensitivity of
the planned next-generation $0\nu2\beta$ experiments~\cite{bbfuture}.
Therefore we focus on studying $\theta_{13}$.

Fig.\fig{theta13} shows the $\Delta\chi^2$
distribution for the predicted $\theta_{13}$,
compared with the present bound from CHOOZ~\cite{CHOOZ} and
SK~\cite{atm} ($\theta_{13}< 10^\circ$ at 90\%{} CL).
Its structure reflects the presence of local minima at different values of $\Delta m^2_{\rm sun}$
in our global fit of solar and KamLAND data, see Fig.\fig{solarfit}a.
With more statistics, KamLAND will be able to measure
the solar oscillation parameters with few $\%$ error~\cite{jhep}.
Long baseline experiments will measure the atmospheric parameters
with few $\%$ error~\cite{LBL}, allowing to predict $\theta_{13}$ with $\sim 10\%$ error.
First-generation long-baseline experiments will
be sensitive to $\theta_{13}\circa{>}0.08$~\cite{LBL}.
The whole predicted range for $\theta_{13}$
can be covered at second-generation experiments, such as JHF~\cite{LBL}.
% The LOW solution predicts $\theta_{13}\approx 0.002$, just below the
% sensitivity of the most optimistic neutrino factory projects~\cite{LBL}.

\medskip

So far we have discussed the predictions for the light-neutrino mass matrix $m_\nu$.
Within this model, oscillation experiments have already fixed it,
except for the CP-violating phase.
Future experiments will test the model.
To get information on the heavy neutrino
masses in\eq{txt}, and to test the high-energy part of the model,
we need an additional input from leptogenesis.

\begin{figure*}[t]
 $$\includegraphics[width=7cm]{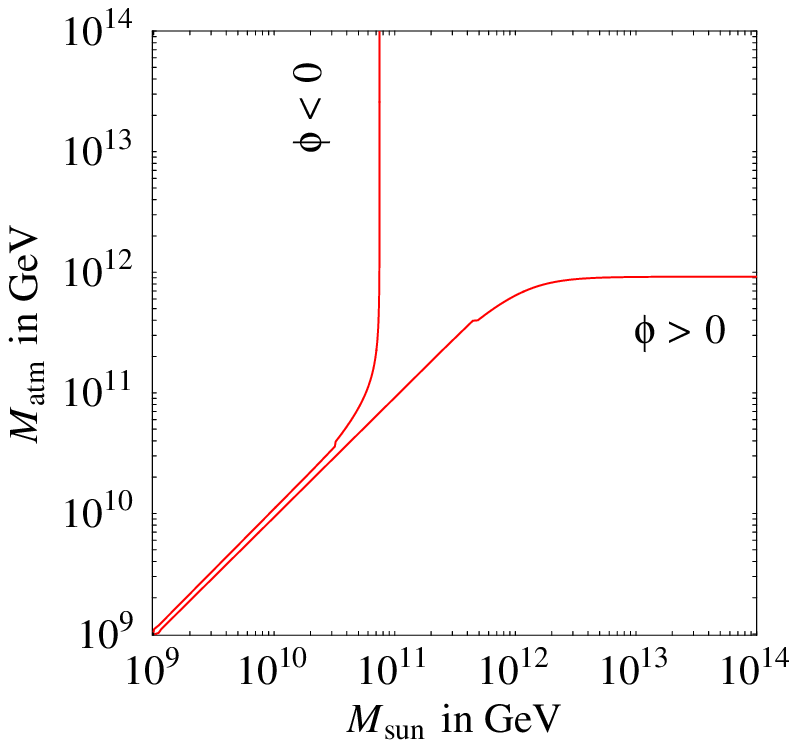}\hspace{1cm}
 \includegraphics[width=7cm]{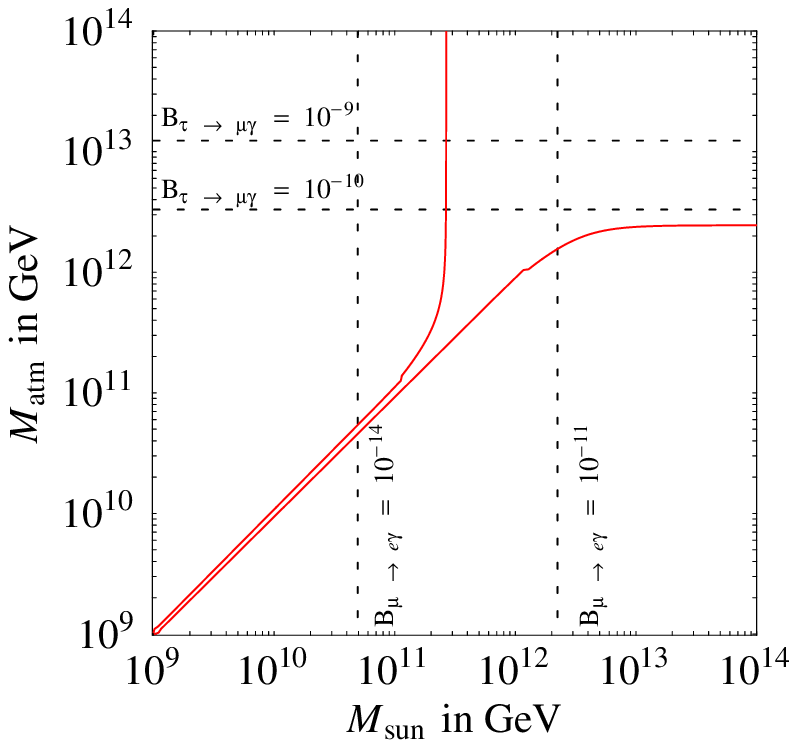}$$
\caption{\em
Heavy neutrino masses $M_{\rm sun}$ and $M_{\rm atm}$ which
imply successful leptogenesis for $|\sin \phi| = 1$ 
and for LMA  best-fit oscillations.
The left (right) plot refers to the non-supersymmetric (supersymmetric) minimal see-saw model.
In the supersymmetric case we also show the contour lines of
{\rm BR}$(\mu\to e \gamma)$ and {\rm BR}$(\tau\to \mu\gamma)$ assuming
$m_0 = 100\GeV$, $M_{1/2}= 150\GeV$, $A_0=0$, $\tan\beta=10$.
\label{fig:MsMa}}
\end{figure*}

\subsubsection*{\em Leptogenesis}
%Assuming that the mass difference between the two heavy neutrinos
%is much larger than their widths,
The decays of the lightest
right-handed neutrino,
$$N_1 = N_{\rm sun}\quad\hbox{or}\quad N_{\rm atm},$$
generate a lepton asymmetry only in
$L_\mu$ (see\eq{txt}).
The generated lepton asymmetry is then converted into a
baryon asymmetry by sphalerons~\cite{sphaleron}.
The baryon-to-entropy ratio in the
non-supersymmetric model is given by
\begin{equation}\label{eq:nBs}
\frac{n_B}{s} = (0.85\pm 0.15)\times 10^{-10} =
-\frac{3\epsilon\eta}{2183},
\end{equation}
where $\epsilon$ is the CP-asymmetry in $N_1$ decays
and $\eta < 1$ is an efficiency factor, determined by
solving the relevant set of Boltzmann equations~\cite{sm}.

In Fig.\fig{MsMa}a we show the iso-curves of the predicted $n_B/s$
in the ($M_{\rm sun}, M_{\rm atm}$) plane
assuming the best-fit values of oscillation parameters.
Unless the heavy neutrinos are extremely degenerate, which we regard as
a fine tuning, Fig.\fig{MsMa} implies that the $N_1$ Yukawa couplings
are sufficiently large that $N_1$ quickly
reaches the thermal abundance and washes out the lepton asymmetry
eventually generated by the heavier singlet neutrino.

The main features of leptogenesis in this model
can be understood by simple analytic approximations as follows.
\begin{itemize}
\item If $M_{\rm sun} \ll M_{\rm atm}$,
$$\epsilon =
\frac{3}{16\pi} \frac{m_{\rm atm} M_{\rm sun}}{v^2}
\frac{s_{23}^2}{1+c_{23}^2 t_{12}^2} \sin\phi .$$
For $M_{1}\ll 10^{14}\GeV$ only $\Delta L=1$ washout scatterings contribute to
the efficiency factor, and $\eta$ is approximately given by~\cite{sm}\footnote{Fig.\fig{MsMa}
is obtained from an accurate numerical computation, performed along the lines of~\cite{sm}.
We include important thermal corrections to the Higgs mass $m_H$ and use
the values of top Yukawa coupling $\lambda_t$, $\tilde{m}$ and $m_{\rm atm}$ renormalized at the $N_1$ mass
($\Delta L = 1$ washout scatterings depend on $m_H$ and $\lambda_t$).
% These factors have been neglected in~\cite{sm} and give ${\cal O}(1)$ corrections.
}
$$\eta \approx 1.5~10^{-4}\eV/\tilde{m},$$
where the effective mass $\tilde{m}$ is given only in terms of the $L_\mu$
interactions in\eq{txt}.
The texture predicts at the best-fit point
$$ \tilde{m} =
m_{\rm sun} \frac{c_{12}^2}{c_{23}^2} \approx 0.01\eV. $$
Thus $\eta\sim 0.01$.
The observed baryon asymmetry is obtained for $\phi<0$
and
$M_{\rm sun}\approx 10^{11}\GeV/|\sin\phi|$ independently of $M_{\rm atm}$.

\item If $M_{\rm atm} \ll M_{\rm sun}$,
\begin{eqnarray*}
\epsilon &=& -\frac{3}{16\pi} \frac{m_{\rm sun}
M_{\rm atm}}{v^2}\frac{t_{23}^2}{1+t_{12}^2} \sin \phi,\\
\tilde{m} &=& s^2_{\rm atm} m_{\rm atm} \approx 0.03\eV,
\end{eqnarray*}
thus $\eta \sim 0.003$.
The observed baryon asymmetry is obtained for $\phi>0$ and
$M_{\rm atm}\approx 10^{12}\GeV/|\sin\phi|$ independently
of $M_{\rm sun}$.

\item If $M_{\rm atm} \approx M_{\rm sun}$ the
CP-asymmetry is enhanced \cite{p} by $1/|M_{\rm atm}-M_{\rm sun}|$  and
reaches a maximum $\epsilon\sim 1$ when the mass
difference is comparable to the decay widths.
The observed baryon asymmetry can be obtained for a large range of
relatively low heavy neutrino masses.
Its sign still depends on which singlet neutrino is heavier,
and it does not fix the sign of CP-violation in oscillations.

\end{itemize}
To summarize,
Fig.\fig{MsMa} implies that we need to know both the sign of
$\phi$ and the sign of the baryon asymmetry to determine the discrete
ambiguity of the model: the mass ordering of the heavy neutrinos.
For hierarchical heavy neutrinos leptogenesis determines the mass of the
lightest one, but does not test the model.

\subsubsection*{\em Supersymmetry and lepton flavour violation}
If nature is supersymmetric,
it could be possible to fix and test the high-energy part of the model.
For leptogenesis and neutrino masses
the presence of supersymmetry  changes only few
${\cal O}(1)$ coefficients:
(i) the vacuum expectation value $v$ is replaced by $v\sin\beta$;
(ii) the CP-asymmetry $\epsilon$ becomes 2 times larger
when $M_{\rm sun}$ and $M_{\rm atm}$ are hierarchical~\cite{vissani}
(iii) numerically eq.\eq{nBs} remains practically unchanged since
the number of model degrees of freedom is about doubled;
(iv) washout becomes more efficient~\cite{Plum}:
$$\eta\approx0.3~10^{-4}\eV/\tilde{m}.$$
The final result is shown in Fig.\fig{MsMa}b which differs from
 Fig.\fig{MsMa}a by a small factor.
Both for $\phi>0$ and $\phi <0$
we observe a potential conflict between obtaining a
successful thermal leptogenesis and avoiding overproduction
of gravitinos~\cite{gravitino} in this model.
If gravitinos do exists,
they either must be heavier than $m_{\tilde G}>10\TeV$ in order to
allow the mass-scales of Fig.\fig{MsMa}b,
or, for $m_{\tilde G}\sim 1\TeV$, one must have $M_1 < 10^8\GeV$.
The last condition is satisfied
only when $M_{\rm sun}$ and $M_{\rm atm}$ are almost degenerate.

\medskip

In supersymmetric extensions of the see-saw model,
the renormalization effects due to the neutrino Yukawa couplings
imprint lepton flavour violation in the slepton masses~\cite{bm}.
Assuming that soft terms are universal at the unification scale
(a hypothesis that collider experiments can partly test),
in a generic see-saw model
$$
\mathscr{W}=\mathscr{W}_{\rm MSSM} +
\frac{M_{ij}}{2}N_iN_j+\lambda_N^{ij}\,N^i L^j H_{\rm u},
$$
the correction to the $3\times 3$ mass matrix of
left-handed sleptons is given by
$$
\mb{m}^2_{\tilde{L}} = m_0^2-\frac{1}{(16\pi)^2} (3m_0^2 + A_0^2)
\mb{\lambda}_N^\dagger
\ln(\frac{M_{\rm GUT}^2}{\mb{MM}^\dagger})
\mb{\lambda}_N+\cdots. $$
In general see-saw models the presence of too many uncontrollable
neutrino parameters does not allow to make real predictions on lepton flavour violation (LFV).
The present model allows us to compute the
$\mu\to e \gamma$ and $\tau\to \mu\gamma$ rates~\cite{hisano}
(and related LFV processes~\cite{der})
in terms of the two high-energy parameters $M_{\rm sun}$ and $M_{\rm atm}$.
Assuming that thermal leptogenesis generates the observed
baryon asymmetry, we get predictions more sharp than what
suggested by a na\"{\i}ve counting of the number of free parameters.
Barring the case of almost degenerate singlet neutrinos
$M_{\rm sun}\approx M_{\rm atm}$
(where only the ratio
\begin{eqnarray}
\frac{\BR(\mu\to e\gamma)}{\BR(\tau\to \mu\gamma)} &=&
\frac{m_\mu^5 \tau_\mu}{m_\tau^5 \tau_\tau}
\frac{\Delta m^2_{\rm sol}}{\Delta m^2_{\rm atm}}
\frac{\sin^2 2\theta_{12}}{\sin^2 2\theta_{23}\cos^2\theta_{23}}
\nonumber \\
&\approx&
0.2
\label{eq:brrate}
\end{eqnarray}
can be predicted),
leptogenesis fixes the mass  of the lightest singlet neutrino
allowing to compute its Yukawa couplings, and consequently, the LFV rates
that it induces.
The predictions depend on the sign of the CP-violating phase $\phi$
measurable in oscillations.
\begin{itemize}
\item
If $\phi <0$, $N_1$ is $N_{\rm sun}$, BR$(\mu\to e \gamma)$ can be predicted
while BR$(\tau\to \mu \gamma)$ remains a function of a single unknown
parameter, $M_{\rm atm}$.
Since $M_{\rm atm}> M_{\rm sun}$ the model also predicts a {\it lower bound}
on BR$(\tau\to \mu \gamma)$.

\item
If instead $\phi>0$, $N_1$ is $N_{\rm atm}$,
BR$(\tau\to \mu \gamma)$ can be predicted,
together with a lower bound on BR$(\mu\to e \gamma)$.
The latter is a function of the unknown $M_{\rm sun} > M_{\rm atm}$.
\end{itemize}
As usual, the predicted LFV rates depend on sparticle masses
which can be measured at colliders. Taking into account naturalness
considerations and experimental bounds, we give our numerical examples for
$m_0 = 100\GeV$, $M_{1/2}= 150\GeV$, $A_0=0$ and $\tan\beta=10$.
In Fig.\fig{MsMa}b we show the iso-curves of the LFV processes
for this input, assuming the best-fit oscillation parameters.
The branching ratios are calculated by solving numerically
the renormalization group equations and using exact formul\ae{}
in~\cite{hisano}.
Both BR$(\mu\to e \gamma)$ and BR$(\tau\to \mu \gamma)$ can be in the
reach of future experiments~\cite{Barkov}.
Their behavior is approximately given by
\begin{eqnarray*}
\hbox{BR}(\mu\to e \gamma) &\approx&
2.7\, r~10^{-12}
\bigg(\frac{M_{\rm sun}}{10^{12}\GeV}\bigg)^2 , \\
\hbox{BR}(\tau\to \mu \gamma) &\approx&
1.5\,r~10^{-11}
\bigg(\frac{M_{\rm atm}}{10^{12}\GeV}\bigg)^2 ,
\end{eqnarray*}
where the logarithmic dependence on the heavy masses is neglected and
we have introduced an approximate scaling factor
$$r \approx
 \bigg(\frac{\tan\beta}{10}\bigg)^2
\bigg(\frac{ 150\GeV}{m_{\rm SUSY}} \bigg)^4, $$
($r=1$ at our reference point)
in order to show the dominant dependence on supersymmetric model
parameters.
In particular, the branching ratios decouple as
$1/m_{\rm SUSY}^4$ if sparticles are heavy.
When sparticles masses will be measured, it will be
possible to present more precise predictions.

For hierarchical heavy neutrinos,
for $|\sin\phi| = 1$,
and for the LMA best-fit oscillation parameters,
the predictions are
$$\begin{array}{ll}
\hbox{BR}(\mu\to e\gamma)\approx 2\,r~10^{-13}\cr
\hbox{BR}(\tau\to \mu \gamma)\circa{>} 3\,r~10^{-12}
\end{array}
\qquad\hbox{for}\qquad\phi<0,$$
and
$$\begin{array}{ll}
\hbox{BR}(\tau\to \mu \gamma)\approx 7\,r~10^{-11}\cr
\hbox{BR}(\mu\to e\gamma)\circa{>} r~10^{-11}
\end{array}
\qquad\hbox{for}\qquad\phi>0.$$
These results imply that,
if also $\tau\to \mu \gamma$ is observed for $\phi<0$, or
if $\mu\to e\gamma$ is observed for $\phi>0$, all the model parameters
in\eq{txt} can be entirely determined.

In this model the electron and muon electric dipole moments~\cite{dip}
 and $\tau\to e \gamma$ are  generated at a negligible level.

\subsubsection*{\em Conclusions}

Unlike the general see-saw model~\cite{di}, the most minimal see-saw
model\eq{txt} allows to determine the low energy neutrino mass
matrix entirely from neutrino oscillation experiments.
The model predicts eq.s\eq{preds},\eq{numpred}.
The sign of the oscillation CP-phase,
together with the sign of the baryon asymmetry, fixes the order of the
two heavy neutrino masses. Unless they are almost degenerate, successful
thermal leptogenesis determines the lightest of them.
The supersymmetric version of
the model predicts either BR$(\mu\to e \gamma)$ or BR$(\tau\to \mu \gamma)$,
depending on the sign of $\phi.$ The other process remains a function of
the heavier neutrino mass only, and has a lower bound on its branching ratio.
If the heavy neutrinos are almost degenerate, the model predicts only the ratio
BR$(\mu\to e \gamma)$/BR$(\tau\to \mu \gamma)$ according to\eq{brrate}.
Observation of the LFV processes allows in principle to fix all the model
parameters in\eq{txt}, and to test its high-energy part.

% $$
% \special{color cmyk 0.1 0.1 0. 0.}\bullet~
% \special{color cmyk 0.2 0.2 0. 0.}\bullet~
% \special{color cmyk 0.3 0.3 0. 0.}\bullet~
% \special{color cmyk 0.4 0.4 0. 0.}\bullet~
% \special{color cmyk 0.5 0.5 0. 0.}\bullet~
% \special{color cmyk 0.6 0.6 0. 0.}\bullet~
% \special{color cmyk 0.6 0.6 0. 0.}\bullet~
% \special{color cmyk 0.7 0.7 0. 0.}\bullet~
% \special{color cmyk 0.8 0.8 0. 0.}\bullet~
% \special{color cmyk 0.9 0.9 0. 0.}\bullet~
% \special{color cmyk 0.8 0.8 0. 0.}\bullet~
% \special{color cmyk 0.7 0.7 0. 0.}\bullet~
% \special{color cmyk 0.6 0.6 0. 0.}\bullet~
% \special{color cmyk 0.5 0.5 0. 0.}\bullet~
% \special{color cmyk 0.4 0.4 0. 0.}\bullet~
% \special{color cmyk 0.3 0.3 0. 0.}\bullet~
% \special{color cmyk 0.2 0.2 0. 0.}\bullet~
% \special{color cmyk 0.1 0.1 0. 0.}\bullet
% $$\Black

\paragraph{Acknowledgements.}
We thank R. Barbieri, J. Ellis, A. Romanino and T. Yana\-gida for discussions.
This work is partially supported by EU TMR
contract No.  HPMF-CT-2000-00460 and by ESF grant No. 5135.

\end{document}